\documentclass[12pt]{iopart}

\usepackage[dvips]{graphicx}   
\newcommand{\beqn}{\begin{equation}}
\newcommand{\eeqn}{\end{equation}}
\newcommand{\beqna}{\begin{eqnarray}}
\newcommand{\eeqna}{\end{eqnarray}}

\begin{document}

\title[Multifractal detrended moving average analysis...]{Multifractal detrended moving average analysis of global temperature records}

\author{Provash Mali}

\address{Physics Department, North Bengal University, Siliguri 734013, India}
\ead{provashmali@gmail.com}
\begin{abstract}
Multifractal structure of global monthly mean temperature anomaly time series over the period of 1850--2012 are studied in terms of the multifractal detrended moving average (MFDMA) analysis. We try to address the possible source(s) and the nature of multifractality in the time series data by comparing the results derived from the actual series with those from a set of shuffled and surrogate series. It is seen that the MFDMA method predicts a multifractal structure of the temperature anomaly records that is more or less similar to what was obtained from the multifractal detrended analysis for the same set of data. In our analysis the major contribution of multifractality in the data is found to be due to the long-range temporal correlation among the measurements, however the contribution of a fat-tail distribution function of the variables is not negligible. The existence of long-range correlation is also confirmed by the constancy of the local slopes of the fluctuation function over a sufficient scale intervals. The results of the moving average analysis are found to depend upon the location of the detrending window and tend to the observations of the multifractal detrended analysis for a specific choice of the location of the detrending window.   
\end{abstract}

\section{Introduction}
We know that most of the natural processes are nonlinear which ultimately lead to fractal measurements. The idea of fractal analysis was introduced by B.B. Mandelbrot in the late 1960s \cite{Mand1,Mand2}. Recently with the advancement of computing facility the study of fractal and multifractal systems has gained an extra dimension in the field of nonlinear dynamics. Now-a-days several methods of multifractal analysis have been furnished and, with their own merits and demerits, they are applied to characterize the time series data of different variants. The simplest type of multifractal analysis is based upon the standard partition function multifractal formalism which has been developed for the multifractal characterization of normalized stationary time series \cite{Bara91, Peit92, Barcy01}. Unfortunately, this standard formalism does not give correct results for nonstationary time series that possess trends or that cannot be normalized. An improved multifractal formalism, namely the wavelet transform modulus maxima (WTMM) method, has been developed in the early 1990s \cite{Muzy91, Muzy94, Arne95, Ivan99, Amar01, Silc01}. The WTMM method is based on wavelet analysis and hence involves tracing the maxima lines in the continuous wavelet transform over all scales. Latter in 2002 Kantelhardt et al. \cite{Kant02} have developed an alternative approach by generalizing the detrended fluctuation analysis (DFA) \cite{Peng94, Ossa94}, which has already been efficiently used in atmospheric data analysis \cite{Eichner03,Maraun04,Varotsos05,Varo13}. The multifractal version of DFA (MFDFA) does not require the modulus maxima procedure, and hence does not involve more effort in computer programming than the conventional DFA. Another important usefulness of MFDFA is that, it is not affected by the underlying trends of the data. Probably because of the simplicity and efficiency of MFDFA, the spectrum of applicability of the method covers all the possible fields of time series analysis within just a decade of its introduction. Usually, the empirical time series data are mostly affected by non-stationarities which have to be well distinguished from the intrinsic fluctuations of the record, in order to find out the correct scaling behavior of the data. Unfortunately, very often we do not know the underlying trends in the data and even worse we do not know the scales of the underlying trends. Recently a novel technique, known as the multifractal detrended moving average (MFDMA) method, has been proposed by Gu and Zhou \cite{Gu10} for the multifractal characterization of time series data. The MFDMA method is a generalization of the DMA method of Alessio et al. \cite{Ales02}, where the local trends of the analyzing signal is filtered out (detrended) by subtracting the local means. The MFDMA method can easily describe the multifractal nature of non-stationary series without any assumption. So far the MFDMA method is used in a limited area of time series analysis \cite{Schu11, Wang11,Ruan11, Shao12, Zhou12, Wang14}, though the empirical studies suggest that under certain circumstances the performance of MFDMA is slightly better than MFDFA \cite{Ruan11,Shao12,Zhou12}.

The global temperature is a crucial thermodynamic parameter of the atmosphere and its rising trend, as illustrated in Fig.\,\ref{fig:series}(a), has become a global issue of climatic research \cite{Eichner03,Maraun04,Varotsos05,Varo13,Kosc98,Weber01,Efst13,Mali14a}. Moreover, the variation of daily and/or monthly mean temperature measured over their smooth average is found to be so random that the data always require an extra attention, in order to characterize the underlying dynamics. In some early analysis of daily maximum temperature fluctuations from their average values \cite{Kosc98} it has been claimed that the fluctuating pattern follows a monofractal scaling relation with time lags, that means a single parameter, called scaling exponent, is enough to interpret a sequence of observations. Latter, the study of Weber and Talkner \cite{Weber01} has revealed that the value of the exponent depends upon the altitude of the meteorological station. In several occasions the DFA method is applied to analyze the temperature record data of different variant \cite{Eichner03,Maraun04,Varo13}. The ultimate finding of these analysis is that, the fluctuation functions derived from a temperature record series of any form is found to follow a power-law type of scaling relation with time lags, and it was interpreted as an effect of long-memory process. However, in Refs. \cite{Maraun04,Varo13} it has been demonstrated that the power-law scaling of the fluctuation functions cannot be regarded {\it a priori}, but it should be established in conjunction with the investigation of the local slopes of the log-log plots of the fluctuation functions. According to \cite{Maraun04}, the comparison of a long-memory process with a short-memory model does not specify the existence of long-range correlations from the application of DFA on a finite data set, and hence scaling cannot be concluded from a log-log straight line fit to the fluctuation function. Recently, we analyze the global monthly mean temperature anomaly time series data in terms of the MFDFA method \cite{Mali14a}. In this analysis we find that the time series records exhibit a rich multifractal structure which originates from two possible sources, namely long-range temporal correlation and fat-tailed probability function of the values. Further, we find that the time series can be more or less described by a computer generated series based on the generalized binomial multifractal algorithm \cite{Kant03}.

The present analysis attempts to investigate the existence of intrinsic scaling properties of the global temperature anomaly records over the period of 1850--2012 \cite{Data} by using the MFDMA method \cite{Gu10}. We establish the power-law scaling of the fluctuation functions from the constancy nature of the local slopes of the log-log plots of fluctuation functions. The investigation shows that the power-law scaling, as it has also been observed in \cite{Mali14a}, is an outcome of long-range correlations of the time series records. The MFDMA results are supplemented by the autocorrelation function analysis, which provides a preliminary idea about the correlation pattern present in the data. The paper is organized as follows: in section \ref{Data} we present the data characteristics. In section \ref{Methods} we provide the details of our analysis, where under two different subsections the aspects of the autocorrelation analysis and MFDMA analysis are discussed. We conclude the paper in section \ref{Conclusions}.

\section{Data}
\label{Data}
\begin{figure*}[t]
\centering
\includegraphics[width=\textwidth]{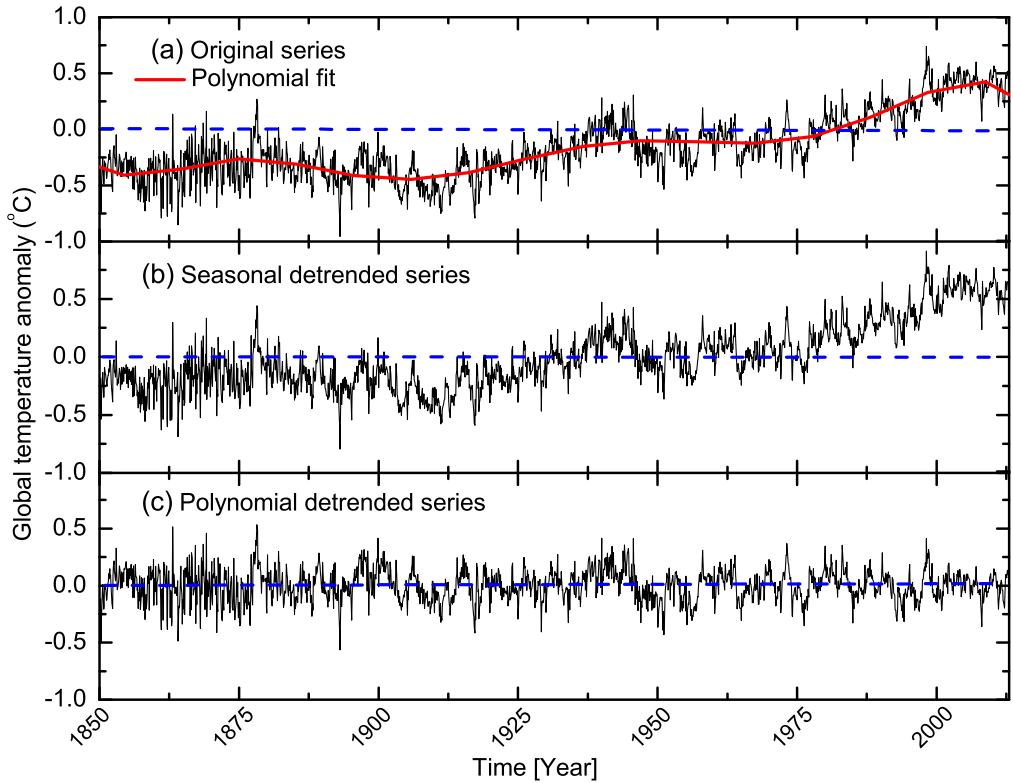}
\caption{Global monthly mean temperature anomaly time series from 1850 to 2012. The anomalies are relative to the mean of the reference period of 1961-1990. (a) The original series with a background trend (red line) best determined by a polynomial of degree 7, (b) the seasonal detrended series (see the text for details) and (c) the polynomial detrended (residue) series corresponding to the series shown in (a). The dotted line represents the $0^{o}$C reference level of the respective series.}
\label{fig:series}
\end{figure*}
\begin{figure}
\centering
\includegraphics[width=0.75\textwidth]{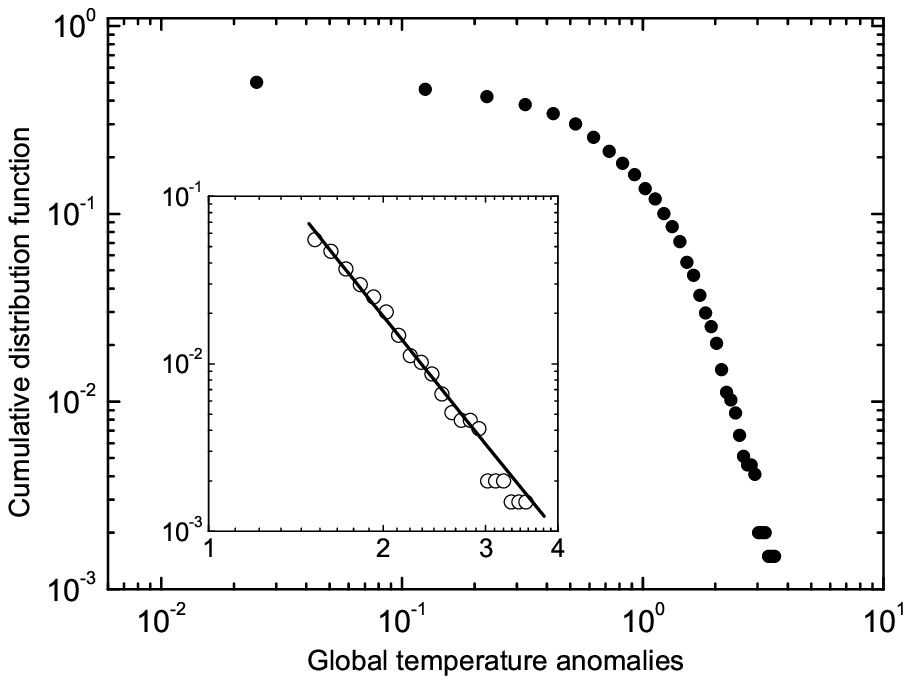}
\caption{Cumulative distribution function for the polynomial detrended temperature anomaly time series. The inset implies that the tail region of the distribution function can be well fitted to a power-law formula with a tail exponent $\alpha_{\rm tail} \approx 4$, indicating the distribution is a fat-tailed one.}
\label{fig:cdf}
\end{figure}
The temperature anomaly time series data used here is taken from the database of the Climatic Research Unit, University of East Anglia \cite{Data}. The global monthly mean temperature anomaly time series from 1850 to 2012 are shown in Figure\,\ref{fig:series}. Note that the anomalies are relative to the mean over the reference period of 1961--1990. Apparently the original series having an overall upward trend contains nonstationarities. According to Environmental Protection Agency (EPA) reports, the earth's temperature has increased by 0.8$^{\circ}$C over the past century and more than half of the increase has happened in the last 25 years. Though apparently it is not visible from the diagram [Figure \ref{fig:series}(a)] but the time series data contains a periodic seasonal trend. In order to eliminate the seasonal trend, we calculate the departures $\mathcal T_i = t_i - \overline{t}_i$ from the mean monthly record $\overline{t}_i$. The monthly mean $\overline{t}_i$ is calculated for each calendar month $i$, e.g. January, by averaging over all the 162 years in the records. The seasonal detrended series illustrated in Figure\,\ref{fig:series}(b) also shows more or less similar upward trend as the original series does. Therefore, the long-term trend (appears as a periodic one) is filtered out by subtracting the smooth background--the best fitted polynomial to the original series. The smooth (red) line in diagram \ref{fig:series}(a) represents the polynomial of degree 7 which gives the best figure of merit. The polynomial detrended anomalies are shown in Figure\,\ref{fig:series}(c), where the long-term periodic trend is no longer present. The subsequent analysis is carried out by using the polynomial detrended anomaly values \ref{fig:series}(c), although in the text it is said to be the original series. In order to specify the statistics of the series variables, we construct the cumulative distribution function (CDF) for the (polynomial detrended) series variables. The CDF is shown in Figure\,\ref{fig:cdf}. The inset in the diagram magnifies the tail region of the CDF. The tail exponent $\alpha_{\rm tail}$ is evaluated by a power-law regression that gives $\alpha_{\rm tail}\approx 4$. As we know that the limit $\alpha_{\rm tail} \sim 3$ is taken as the onset of a fat-tailed distribution. Accordingly, the underlying probability distribution function for the data is a fat-tailed one, and hence the distribution function may in principle be a source of multifractality.

\section{Analysis and Results}
\label{Methods}
\subsection{Autocorrelation function}
Autocorrelation function for a time series data provides the correlation between the $i$th measurement with that of the $(i+s)$th one for different values of time lag $s$. Consider a time series $\{x_i:i=1,2,\cdots,N\}$, here the index $i$ corresponds to the time of a measurement $x_i$. In order to remove the constant offset of the series (if any), the mean of the series $\left<x\right>=(1/N)\sum_{i=1}^N x_i$ is subtracted: $\bar{x}_i=x_i-\left<x\right>$. Then the auto-covariance between any two $\bar{x}$'s separated by $s$ steps (or lag) is defined as
\begin{equation}
C'(s) = \left<\bar{x}_i \bar{x}_{i+s}\right> = \frac{1}{N-s}\sum_{i=1}^{N-s}\bar{x}_i \bar{x}_{i+s}. 
\end{equation}
When the above $C'(s)$-function is normalized by the variance $\left<\bar{x}^2_i\right>$, the function is called the autocorrelation function $C(s)$. If the series $\{x_i\}$ are uncorrelated, $C(s)$ is zero for any $s>0$. The $\{x_i\}$s are said to be short-range correlated, if $C(s)$ declines exponentially: $C(s) \propto \exp(-s/s_0)$ for $s \to \infty$. On the other hand, for a long-range correlated series, $C(s)$ declines as a power-law: $C(s) \propto s^{-\gamma}$ for $s\to \infty$ with exponent $0 < \gamma <1$. A direct calculation of $C(s)$ is usually not appropriate due to the noise superimposed on the series $x_i$ and due to the underlying trends of some unknown origin, and hence the exponent $\gamma$ is extracted indirectly. In this analysis we employ the MFDMA technique to capture the nature of correlation present in the temperature anomaly records. However, for a time series analysis the autocorrelation function analysis is always appreciable, as it may provide an elementary idea about the type of correlation in the data. Note that the pattern of correlation for a stationary time series may also be studied in terms of the so-called power spectrum $E(f)$ at frequency $f$: $E(f) \sim f^{\beta}$. For stationary time series the exponent $\beta$ is related to $\gamma$ through $\gamma=1-\beta$. 
\begin{figure}
\centering
\includegraphics[width=0.75\textwidth]{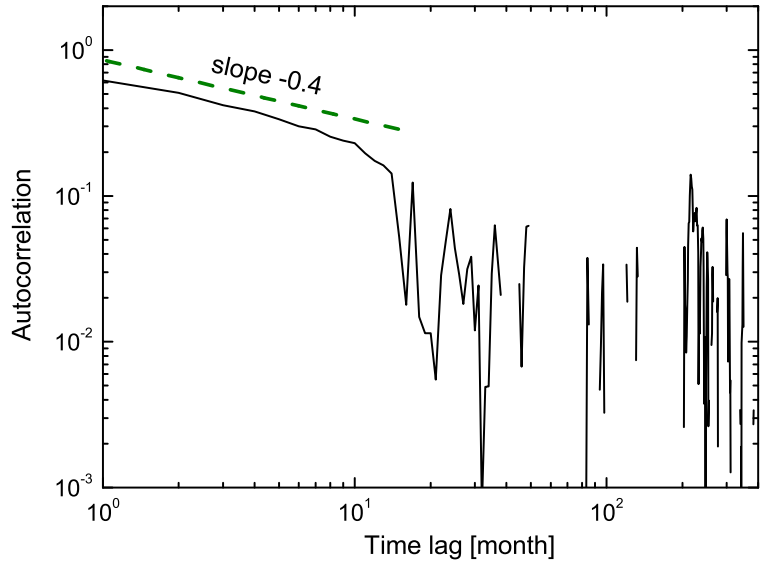}
\caption{Autocorrelation function for the global monthly mean temperature anomalies. The dashed straight line having slope $-0.4$ is for the visual reference only.}
\label{fig:auto}
\end{figure}

In Figure\,\ref{fig:auto} we illustrate the autocorrelation function $C(s)$ with time lags $s$ for the temperature record data (actually for the polynomial detrended sequences, Figure\,\ref{fig:series}(c)). The autocorrelation function is found to follow the power-law scaling $C(s) \sim s^{\gamma}$ in the scale interval $1 \leq s \leq 12$ with exponent $\gamma \approx -0.4$. The exact trend for a $C(s)$-function with exponent $-0.4$ is shown in the diagram by the dashed line. The discontinuities and huge fluctuations at large $s$ might be due to the limited statistics of the data and/or it might be an indication of the fact that the correlation does not hold at large lag. The hypothesis of studying autocorrelation function at scale $s \to \infty$ is an ideal concept only. For observational data some limitations of measurement always restrict the analysis within a time domain; the minimum scale of which is specified by the sampling interval $\Delta \tau$, whereas the maximum scale is determined by the length of the series $N$.  It is not certain if an autocorrelation function having a maximum time lag of about 12 months can be used to identify the nature of correlation in a time series data, specially when the long-range correlation is suspected. However, the estimated $\gamma$ value ambiguously gives a preliminary indication of the existence of long-range correlation in the temperature anomaly series studied here.

\subsection{Multifractal detrended moving average analysis}  
The MFDMA methodology is well described in Ref. \cite{Gu10}. For the shake of completeness, we briefly outline the procedure step-by-step in the following subsection, however we do not claim the originality of Ref. \cite{Gu10}.

Let $\{x_i:~i=1,~2, \cdots,N\}$ be a time series of length $N$. The MFDMA procedure consists of the following few steps:
\begin{enumerate}
\item Construct a sequence of cumulative sums
\beqn
y(i) = \sum_{k=1}^{i} x_k, ~~ i = 1, 2, \cdots, N.
\label{eq:cumsum}
\eeqn
In the subsequent steps the above sequence is considered as the signal.

\item  Calculate the moving average function $\widetilde y(i)$ in a moving window of size $n$
\beqn
\widetilde y(i) = \frac{1}{n} \sum_{k=- \lfloor (n-1)\theta \rfloor}^{\lceil (n-1)(1-\theta)\rceil} y(i-k),
\eeqn
where $\lfloor \xi \rfloor $ is the largest integer not larger than $\xi$ and $\lceil \xi \rceil $ is the smallest integer not smaller than $\xi$. Here $\theta$ is a parameter $\in [0,~1]$ that specifies the position of the moving window. In general the moving average function includes $\lceil (n-1)(1-\theta)\rceil$ data points in the past and $\lfloor (n-1) \theta \rfloor$ data points in the future. We consider three different values of $\theta = 0,$ 0.5 and 1. For $\theta =0$ the moving average function $\widetilde y(i)$ is calculated over all the past $(n-1)$ data points of the signal, and hence it refers to the backward moving average. In the case of $\theta=0.5$ the function $\widetilde y(i)$ includes half past and half future information in each window, and it is said to be the central moving average. In the third option $\theta=1$, where the moving average function $\widetilde y(i)$ is calculated over all the $(n-1)$ data points in the future, is known as the forward moving average.

\item Detrende the sequences $y(i)$ by subtracting the moving average function $\widetilde y(i)$ and obtain the residue series
\beqn
e(i) = y(i) - \widetilde y(i),
\label{eq:resi}
\eeqn
where $i$ satisfy the criterion: $n- \lfloor(n-1)\theta\rfloor \leq i \leq N - \lfloor(n-1) \theta\rfloor$.

\item Divide the residue series $e(i)$ into $N_s = \lfloor N/n -1 \rfloor$ non-overlapping segments of equal length $n$. Let the segments are denoted by $e_v$ so that $e_v(i) = e(l+i)$ for $1 \leq i \leq n$ with $l=(v-1) n$. For an arbitrary segment $v$ the mean-square fluctuation function $F^2_v(n)$ is calculated as a function of $n$ through 
\beqn
F_v^2(n) = \frac{1}{n} \sum_{i=1}^n \{e_v(i)\}^2.
\eeqn

\item The $q$th order overall fluctuation function $F_q(n)$ is then determined as 
\beqna
F_q(n) &=& \left\{ \frac{1}{N_n} \sum_{v=1}^{N_n} [F^2_v(n)]^{q/2} \right\}^{1/q} ~\mbox{for all } q \neq 0,  
\label{eq:Fq1}\\
F_q(n) &=& \exp\left\{ \frac{1}{2N_n} \sum_{v=1}^{N_n} \ln [F^2_v(n)] \right\} ~\mbox{for } q = 0.
\label{eq:Fq2}
\eeqna

\item The scaling behavior of $F_q(n)$ is examined for several different values of the exponent $q$. For a multifractal series $F_q(n)$ for large values of $n$ would follow a power-law type of scaling relation, such as
\beqn
F_q(n) \sim n^{h(q)},
\label{eq:Scaling}
\eeqn
and the exponent $h(q)$ would be a function of $q$. 
\end{enumerate}

The exponent $h(q)$, known as the generalized Hurst exponent, is an important parameter for a multifractal analysis. For $q=2$ the $h(q)$ exponent is related to the correlation exponent $\gamma$ and the power-spectrum exponent $\beta$ through $h(2) = 1 - \gamma/2 = (1+\beta)/2$. For stationary time series such as the fGn (fractional Gaussian noise), $h(q=2)=H$--the well known Hurst exponent, and also the exponent satisfies the criterion $0 < h(q=2) < 1.0$ \cite{Chianka05}. In the case of a non-stationary signal, e.g. the fBm (fractional Brownian motion), $h(q=2)= H + 1$ and for such signals $h(q=2) > 1$ \cite{Peng94, Mova06}. For a monofractal series with a compact support, on the other hand, $h(q)$ is independent of $q$. Knowing $h(q)$ one can easily derive the multifractal scaling exponents $\tau(q)$ through
\beqn
\tau(q) = q h(q) - 1.
\eeqn
A nonlinear $\tau(q)$ spectrum signals the existence of multifractal nature of the data. For a monofractal process $\tau(q)$ is a linear function of $q$. The generalized multifractal dimensions is given as
\beqn
D(q) \equiv \frac{\tau(q)}{q-1} = \frac{qh(q) - 1}{q-1}. 
\eeqn
For a monofractal time series though $h(q)$ is independent of $q$, $D(q)$ depends on $q$. Another important variable of a multifractal analysis is the multifractal singularity spectrum $f(\alpha)$, which is related to $\tau(q)$ via a Legendre transformation \cite{Halsey86,Pei92}: $\alpha = \partial \tau(q) / \partial q$. The multifractal spectrum $f(\alpha)$ is defined as 
\beqn
f(\alpha) = q \alpha- \tau(q).
\eeqn
Here $\alpha$ is the singularity strength or the H\"{o}lder exponent. For a monofractal structure only one $\alpha$ exponent is expected to describe the system and the corresponding $f(\alpha)$-spectrum would appear as a delta function. On the other hand, for a multifractal structure a spectrum of $\alpha$s is observed which leads to the existence of a $f(\alpha)$ spectrum.

The parameters $f(\alpha)$ and $\tau(q)$ can also be used to provide a thermodynamical description of a random chaotic system \cite{Bohr87a,Bohr87b}. In this approach $\tau(q)$ is analogous to the free energy and its Legendre transformation $\alpha$ is analogous to the entropy of the system. In general, the function $\tau(q)$ exhibits two different realms which are separated by a ``critical value'' of $q = q_{c}<0$. In the thermodynamical interpretation of multifractality this is called two distinct phases of the system and turning $q$ over its critical value is said to a ``phase transition''. Here the parameter $q_c$ plays the role of the inverse of transition temperature.

It is to be noted that the moving average method shares many ideas with the detrended fluctuation analysis, but an added advantage in the former method probably makes it more sophisticated over the latter one. The advantage in MFDMA analysis is that, it gives us the freedom of choosing the location of detrending window with respect to the measurement to be detrended. For instance, for a computer generated series based on the dynamical random cascade model with log-Poison distribution \cite{Kant02} the MFDFA method is not found to be very sensitive. For this particular series the MFDFA estimated values of $h(q)$ for $q<0$ grossly deviate from their analytic values but the MFDMA estimated values of the parameter offer a reasonable agreement with their analytic values. Both the methods are affected by statistical limitations. Note that so far there is no systematic comparative study between these two methods available in the literature, however in a few occasions it is claimed that the MFDMA method is somewhat superior over the other one \cite{Ruan11, Shao12, Wang14}. A reverse observation is also reported in Ref. \cite{Xu05} but it is not on the multifractal form of the methods.

\subsection{Results of the MFDMA analysis} 
\label{Result2}
\begin{figure}
\centering
\includegraphics[width=\textwidth]{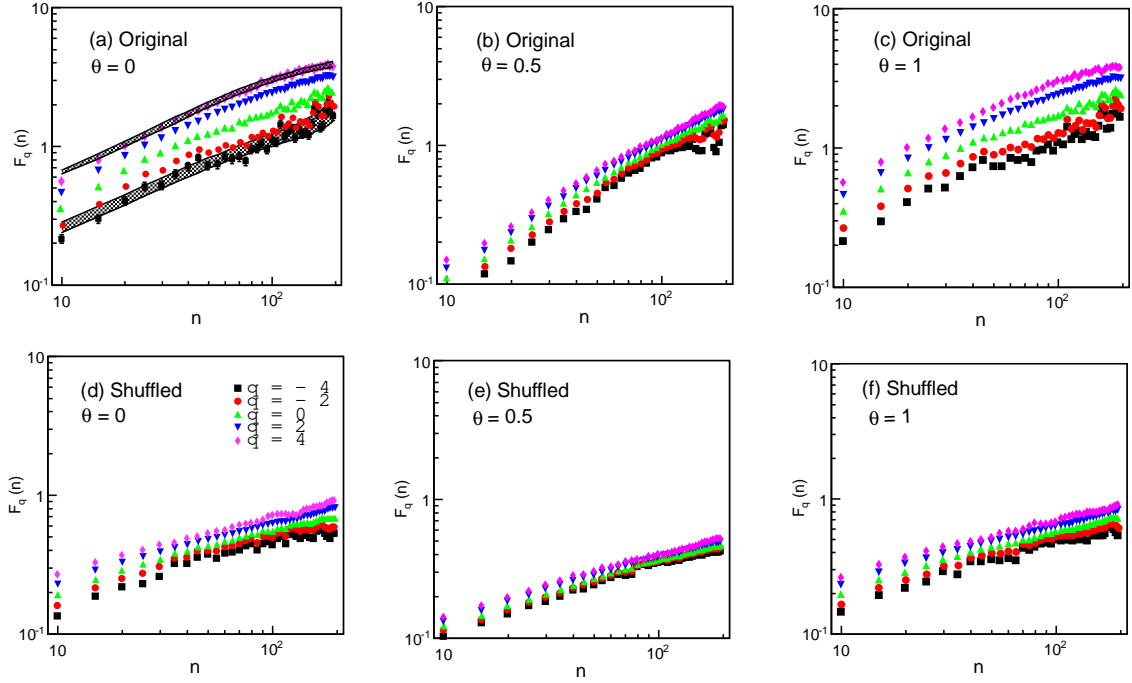}
\caption{Scaling behavior of the MFDMA fluctuation functions $F_q(n)$ for $q=0,\, \pm2,\,\pm4$ for three different choices of $\theta$. The upper panel represents the original series and the lower panel represents an arbitrary shuffled series corresponding to the original one. In diagram (a) the 95\% confidence band of $F_{\pm 4}(n)$ is indicated by the shadowed region.}
\label{fig:Fq}
\end{figure}

We calculate the MFDMA fluctuation functions $F_q(n)$ as a function of window size $n$ (scale parameter) for three different choices of the window parameter $\theta = 0,~0.5$ and 1. The scale parameter $n$ is varied from $10$ to $N/10$ and the exponent $q$ is varied from $-4$ to $+4$ in steps of 0.25. The tail exponent of a CDF $\alpha_{\rm tail}$ usually sets the limits on $q$. In general, for $\alpha_{\rm tail} \geq 3$ the underlying distribution function is classified as a fat-tailed distribution and the exponent $q$ in the interval $\pm 3$ provides the desired information of a fractal measurement. Beyond the specified limits the multifractal variables, such as the exponents $h(q)$, $D(q)$ etc., are expected to show a linear asymptotic behavior, since the limit of the $p$-norm (or H\"older norm) of a vector $\bf x $ of components $\{x_i\}$ as $p$ goes to infinity is the infinity norm, i.e. the supreme of the absolute values of the function ${\bf x}_{\infty} = \max \limits_{1\leq i \leq n}|x_i|$ \cite{Alfio}. As a consequence, the $q$-moment of any variable is rapidly dominated by $\{{\bf x}_{\infty}\}^q$. Corresponding to each of the $\theta$ values the scaling pattern of $F_q(n)$ for $q=0,~\pm 2,~\pm 4$ are shown in Figure\,\ref{fig:Fq}. We also repeat the analysis for a set of 10 randomly shuffled series as well as 10 surrogate series. The lower panel of Figure\,\ref{fig:Fq} illustrates the $F_q(n)$ functions calculated from an arbitrary shuffled series. The scaling of surrogate series generated $F_q$ functions are apparently similar to what is obtained from the shuffled series, and hence they are not pictorially shown. The importance of analyzing the shuffled and surrogate series are discussed below. The statistical error bars for the fluctuation functions are invisibly small in this plot. The 95\% confidence bands are calculated for all the $F_q(n)$ functions but in the figure the confidence band is shown only for $q = \pm 4$. This is because of the clarity of the diagrams. The 95\% confidence bands is estimated by a Monte Carlo technique similar to the one specified in \cite{Maraun04}. In this procedure we generate 100 surrogate series corresponding to the original anomaly series and calculate the $F_q(n)$ functions for all 100 of them. Then for a fixed scale $n$ we obtain the distributions of $F_q$. The distribution function for a known $q$ is then approximated by a Gaussian with mean $\bar{F}_q(n)$ and standard deviation $\sigma_q(n)$ (say). Finally the 95\% confidence band is defined as $\bar{F}_q(n) \pm 1.96 \sigma_{F_q(n)}$.

From Figure \ref{fig:Fq} one can infer that the functions $F_q$ nicely respect the scaling relation (\ref{eq:Scaling}) but mainly in the scale interval $10 \leq n \leq 50$. Above $n \sim 50$ the $F_{q<0}$ functions are highly fluctuating and a nonlinearity is also visible at large scale. One can also notice that the backward ($\theta=0$) and forward ($\theta=1$) moving average schemes result almost similar $F_q$ functions, while the central ($\theta=0.5$) moving average scheme produces slightly stiffer and closely spaced $F_q$s in comparison with the other two schemes. The observations do not require any explanation at this point of our analysis. However, it is clear that the central moving averaging creates a residual series $e(i)$ [Eqn. (\ref{eq:resi})] which is a less correlated one than that produces in the forward/backward moving averaging.            

\begin{figure}
\centering
\includegraphics[width=\textwidth]{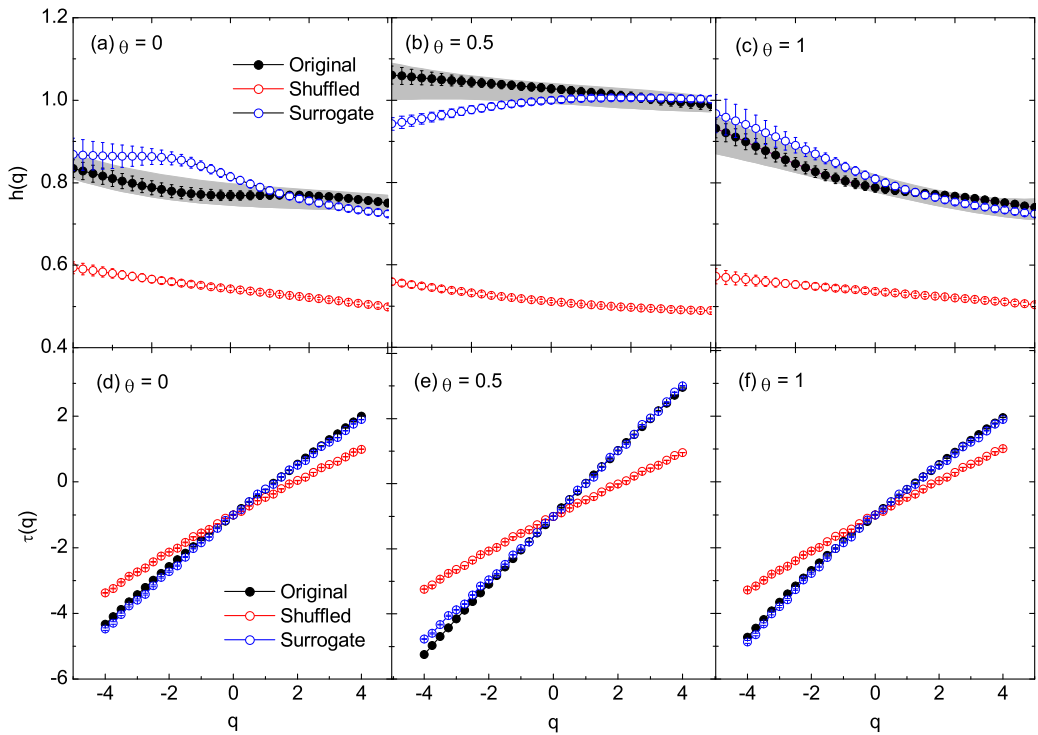}
\caption{Upper panel: Generalized Hurst exponent spectra for the backward ($\theta = 0$), central ($\theta=0.5$) and forward ($\theta=1$) moving averaging schemes. The shadowed regions imply the 95\% confidence bands. Lower panel: Multifractal exponent spectra for the three choices of $\theta$. In all the cases the original series estimated spectra are compared with their respective shuffled and surrogate predictions.}
\label{fig:hq}
\end{figure}
The importance of analyzing a shuffled series for a given empirical time series data is that, a direct comparison between the results obtained from an original series and a shuffled series gives an insight into the nature of multifractality present (if any) in the data \cite{Kant02,Mova06}. A simple way to check if correlations in the data produce any kind of scaling, is to shuffle the data. Random shuffling removes temporal correlation and any scaling remains must be due to the probability distribution of the variables, since the shuffling procedure does not affect the probability distribution function (PDF). If a time series contains multifractality that is stemmed from both correlation and PDF, the corresponding shuffled series will exhibit weaker multifractality than the actual series. The surrogate (also called the phase randomized) series analysis, on the other hand, is a numerical technique of testing nonlinearity in a time series data \cite{Ulenbeck,Thomas}. The aim here is to test whether the dynamics are consistent with some linearly filtered noise or a nonlinear dynamical system. The basic idea of a surrogate data method is to first specify some kind of linear stochastic process that mimics the ``linear properties'' of the original data. If the predictions (statistics) of the original data are significantly different from those of the surrogate series, we may consider the presence of some higher order temporal correlations, that is the presence of dynamic nonlinearities. In this analysis we use the Amplitude-Adjusted Fourier Transform (AAFT) algorithm \cite{Thomas, Theiler}, a well known method, for the surrogate data.

The generalized Hurst exponent $h(q)$ is calculated by fitting a linear function, like
\beqn
\ln F_q(n) = h(q) \ln n + \zeta,
\eeqn
to the $\ln F_q$ versus $\ln n$ data points, but within a limited scale interval: $10 \leq n \leq 50$. Note that the $F_q$ functions are more or less linear and do not possess significant statistical fluctuation within that limits. The $h(q)$ values are plotted against their order number $q$ in Figure \ref{fig:hq} (upper panel) for all three choices of $\theta=0$, 0.5 and 1. In our fitting procedure Pearson's $R^2$ coefficient, measures the goodness of a fit, is found in the interval $0.93 < R^2 < 1$. The $R^2$ values ensure that the fit quality is quite good. The errors associated with $h(q)$ (shown in Figure \ref{fig:hq}) are of statistical origin. The shadowed region describes the $95\%$ confidence bands for the original series estimated $h(q)$ spectrum. The bands are measured as $\bar h(q) \pm 1.96 \sigma_{h(q)}$, where $\bar h(q)$ and $\sigma_{h(q)}$ are the mean and variance of the Gaussion function fitted to the $h(q)$ distribution which is obtained by a similar procedure as mentioned before. The lower panel of Figure \ref{fig:hq} represents the spectra of the multifractal scaling exponent: $\tau(q) = q h(q) - 1$. Since the $\tau(q)$ exponents are directly calculated from the $h(q)$ exponents, we do not put any additional emphasis on this parameter. All the $h(q)$ and $\tau(q)$ spectra are supplemented by their respective shuffled and surrogate counterparts. In order to optimize the randomness of the shuffled/surrogate series generated $h(q)$ (and $\tau(q))$ spectra, an average value of these spectra over 10 independent calculations is considered. The figure reflects the nonlinear nature of the $h(q)$ and $\tau(q)$ spectra. It is seen that for the original and surrogate series the central moving averaging yields a significantly larger values of $h(q)$ and the degree of nonlinearity in the $h(q)$ spectrum is also weaker than the forward and backward moving averaging. According to the theory of multifractals, the $h(q)$ and $\tau(q)$ spectra carry a clear signal of multifractality in the global monthly mean temperature records, but the spectra immensely depend upon the location of the moving window. The forward and the backward moving methods possess some kind of similarity with each other. From this observation one cannot say which detrending window (backward, central or forward) suits better for the time series data analyzed here. For this purpose the results of the detrended moving average analysis have to be systemically compared with that of other known multifractal methods, as well as with various model computations. From Figure\,\ref{fig:hq} it is also seen that the AAFT surrogate series generated spectra, to some extent, take care of their empirical values but the shuffled series generated spectra are underestimated by the corresponding original/surrogate series. Further, the shuffled series calculated $h(q)$ values are all located at $\sim 0.5 - 0.6$ and almost linear in $q$. These observations indicate that the correlation present in the actual series is probably destroyed by the random shuffling, and therefore the shuffled series shows a weak multifractal pattern which is stemmed out of the distribution function (fat-tailed) of the series variables.

\begin{table}
\caption{The second order generalized Hurst exponent $h(q=2)$ values obtained from MFDMA analysis. The estimate of the MFDFA (first order) \cite{Mali14a} is also given for comparison. The errors are statistical only. Pearson's $R^2$ coefficients are quoted under the parenthesis.}
  \label{tab1}
  \begin{center}
    \begin{tabular}{cccccc}
    \hline \hline
    Method && Original & Shuffled & Surrogate \\
    \hline
  MFDMA ($\theta=0$)   && $0.769\pm0.007$  & 0.522$\pm$0.002  & 0.756$\pm$0.008 \\
             && (0.981)          & (0.991)          & (0.943) \\          
  FMDMA ($\theta=0.5$)  && $1.008\pm0.011$    & 0.497$\pm$0.002  & 1.006$\pm$0.007 \\
             && (0.985)           & (0.973)          & (0.933) \\
  FMDMA ($\theta=1$) && $0.767\pm0.007$    & 0.521$\pm$0.002  & 0.753$\pm$0.008 \\
             && (0.945)           & (0.973)          & (0.963) \\
  FMDMA ($\theta=0.7$) && $0.919\pm0.067$    & 0.523$\pm$0.006  & 0.919$\pm$0.017 \\
            && (0.943)           & (0.991)          & (0.993) \\
   MFDFA (Linear)       && $0.907\pm0.011$    & 0.515$\pm$0.007  & 0.767$\pm$0.003 \\
                && (0.931)           & (0.989)          & (0.981) \\
   \hline  
  \end{tabular}
  \end{center}
\end{table}
\begin{figure}
\centering
\includegraphics[width=\textwidth]{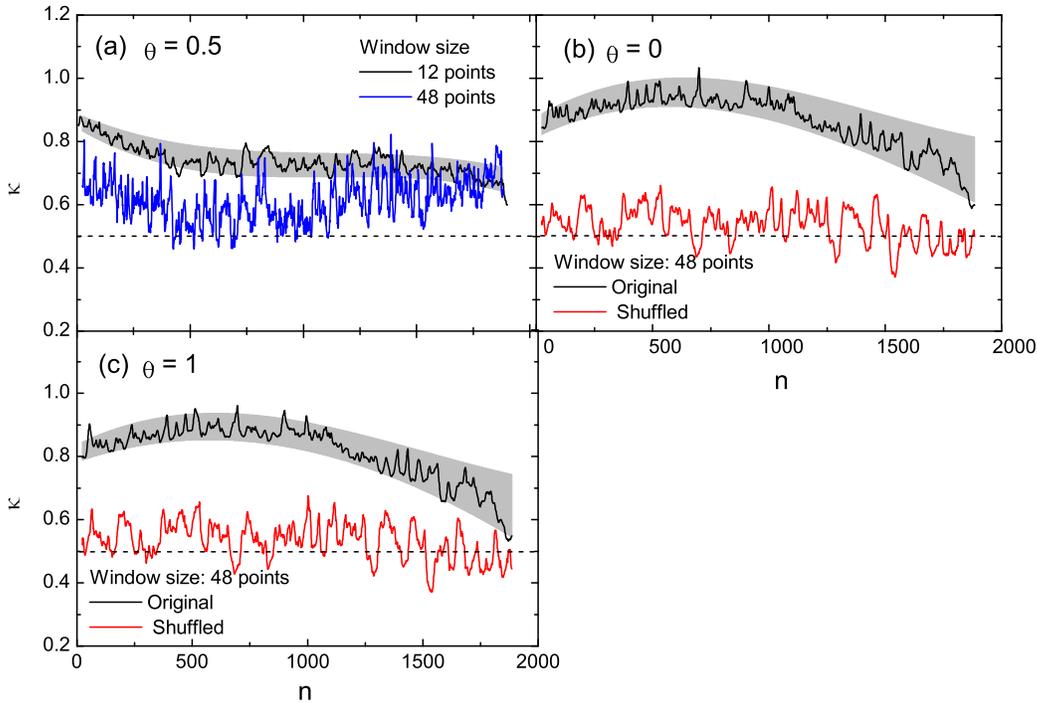}
\caption{Local slopes of the $\ln F_q(n)$ versus $\ln n$ plots (a) obtained from the original series with $\theta=0.5$ for two different values of the window size, (b) prediction of the original series (black) is compared with the corresponding shuffled series (red) for $\theta=0$, and (c) the same as (b) but for $\theta=1.0$. The shadowed band associated with the spectrum represents the $1.96\sigma$ (95\%) confidence limits of that spectrum.}
\label{fig:ls}
\end{figure}
\begin{figure}
\centering
\includegraphics[width=\textwidth]{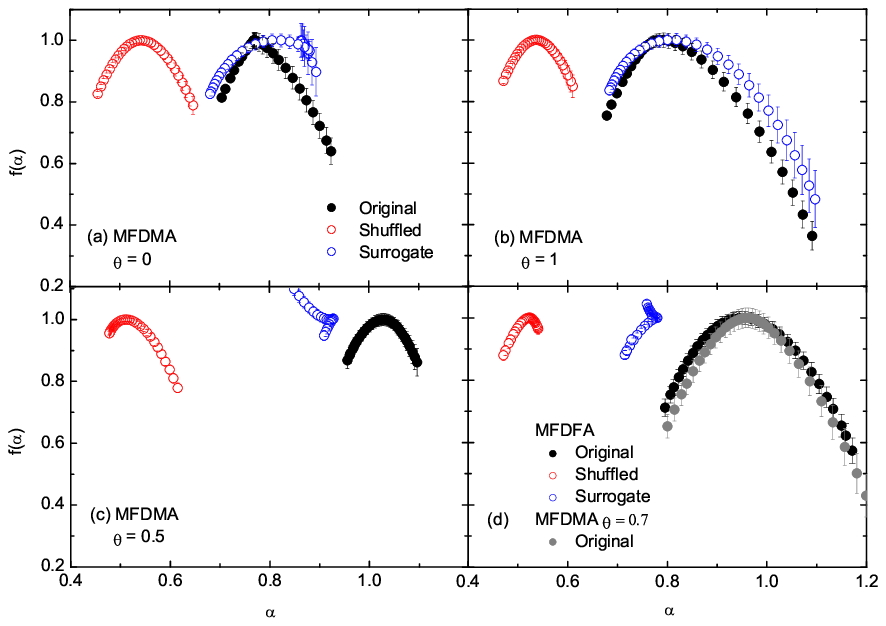}
\caption{(a)--(c) Multifractal spectra, respectively for the backward ($\theta = 0$), central ($\theta=0.5$) and forward ($\theta=1$) moving window. (d) The multifractal spectrum of MFDFA \cite{Mali14a} is compared with that of the MFDMA ($\theta=0.7$) analysis.}
\label{fig:fa}
\end{figure}
The values of second order generalized Hurst exponent $h(q=2)$ obtained for $\theta=0$, 0.5 and 1 are given in Table \ref{tab1}. In addition, we also show the $h(2)$ values for $\theta=0.7$; the reason of which will be discussed latter. The table shows that the original series calculated values of $h(2)$ are very close to their AAFT surrogate series estimation, however the $h(q)$ values obtained from the shuffled series are ($\sim 0.5$) much shorter than their original/surrogate values. The degree of goodness of the measurement of $h(2)$ through straight fit to the $\ln F_2(n)$ versus $\ln n$ data is specified by Pearson's $R^2$ coefficient. The $R^2$ values here are found to be quite satisfactory. It is to be noted that the accuracy of a time series analysis depends upon the length of the series \cite{Zhou12, Alfi06, Bashan08}. Here we find about $5\%$ increase in the $h(2)$ values when the series length $N$ is truncated by 500 points (measurements). The dependence of the confidence band-width on $N$ is found to be more pronounced. For instance, when the $h(2)$ exponent is increased by approximately 5\% because of the truncation of the series by $500$ points, its confidence band becomes wider by more than 12\%.

We can now compare the results of MFDMA analysis with those of the autocorrelation analysis. Recall the autocorrelation exponent value for the data $\gamma\approx 0.4$. This yields $h(2) = 1-\gamma/2 \approx 0.8$. Thus the autocorrelation function estimated value of $h(2)$ roughly match the forward and backward moving average estimated values ($\sim 0.76$). Moreover, for the original series the moving average method with $\theta=0.7$ estimates $h(2)$ value that is very close to the value obtained from the MFDFA (first order) analysis \cite{Mali14a}. From the $h(q)$ exponents/spectra we understood that, in the case of an uncorrelated (AAFT-surrogate) series, where a weak multifractal structure might appear due to the fat-tailed PDF (linear correlations), the two detrended analysis methods differ significantly. Considering the values of $\gamma$ and $h(2)$ exponents, one can argue that the global temperature records behave more or less like a stationary time series for which $0 < h(q=2) < 1.0$ \cite{Chianka05}, though the series is not a stationary one. This implies that autocorrelation function is not a suitable tool to characterize time series data.

As it has been argued in \cite{Maraun04} that the power-law scaling of the detrended fluctuation functions should not be taken as the evidence of long-range correlations. Rather it has to be established from the constancy of local slopes $\kappa$ of the fluctuation functions over a sufficient scale range. Though the extend of the range cannot be defined yet \cite{Avnir98}. Figure \ref{fig:ls} illustrates the local slopes $\kappa$ of the MFDMA fluctuation functions $F_{q=2}(n)$ versus the scale $n$. In diagram (a), where the central moving scheme ($\theta=0.5$) is used, the variation of $\kappa$ with $n$ is shown for two different values of window size: $w = 12$ and 48 months. It is seen that for $w=12$ the $\kappa$ values are very much chaotic, whereas for $w=48$ and onward it possess very little fluctuations and the values are, within the 95\% confidence bands, approximately constant over the scale range $n \simeq 500 - 1400$. Beyond the specified limits of $n$ the local slope values decline very slowly with increasing $n$. In the case of $\theta=0$ (b) and $1$ (c) the constancy interval of $\kappa$ is shifted towards the low $n$ region: $n \leq 1000$. The shuffled series estimated $\kappa$ values are always highly fluctuating at about $\kappa=\kappa_0 \sim 0.55$. The observation supports the possibility of a long-memory process in the data.

Next, we calculate the multifractal singularity spectrum $f(\alpha)$ for the analyzed time series data. The importance of it in connection with a multifractal analysis is that, the parameter itself gives a direct and quantitative measure of the degree of multifractality present in the data. The width and (a)symmetry parameters of the spectrum are closely connected to the chaotic/fractal nature of the data: a wider and asymmetric singularity spectrum roughly imply the time series is more chaotic (rich structure) compared to a series that produces narrower and symmetric singularity spectrum. Also the location of the spectrum gives another important information. For an uncorrelated series the mean of the spectrum is usually spotted at $\alpha \sim 0.5$ but for a long-range correlated series the mean is expected to be shifted at large $\alpha$. In Figure\,\ref{fig:fa} the singularity spectra of our analysis are plotted against the singularity (H\"older) exponent $\alpha$. Separate diagrams are shown for the three choices of $\theta$: (a) backward ($\theta=0$), (b) forward ($\theta=1$) and  (c) central ($\theta=0.5$). In diagram (d) the singularity spectrum of the MFDFA (first order) method \cite{Mali14a} is compared with that of the MFDMA method with $\theta=0.7$. We find that (i) the original series for all the choices of $\theta$ results a stable and wider singularity spectrum, (ii) the surrogate spectrum for $\theta=1$ more or less matches the empirical values, otherwise the surrogate spectra are mostly unstable, (iii) in all the cases the spectra for the shuffled series are located at $\sim 0.5$ and they are narrower than their original series generated counterpart and (iv) the prediction of the MFDMA analysis with $\theta=0.7$ is approximately identical to that of the MFDFA technique. All these observations are related to the fact that the degree of chaoticity/multifractality in the actual series is higher than their shuffled and/or surrogate partner. Once again we observe the effect of random shuffling in the singularity spectra. The weak multifractal effect visible in the shuffled series generated $f(\alpha)$ spectra probably arise from the fat-tailed distribution function of the series values.

At the end of this section we tie up a comparative study between the MFDMA and MFDFA techniques of time series analysis with a reference to the global temperature anomaly time series data analyzed here. Note that by doing so we do not mean that the MFDFA method is a standard method of multifractal time series analysis, though it has been extensively applied on various fields of stochastic data analysis\footnote{The list of references is too long to cite. To get a comprehensive idea follow Refs. \cite{Kant02,Mali14a,Mova06,Gires,Yu14,Mali14,Mali15} and the references therein.}. In this analysis we try to adjust the location of the detrending window, in order to minimize the deviation between the $f(\alpha)$ spectra of MFDMA and MFDFA (first order) \cite{Mali14a} methods. In this process we find the best match at $\theta=0.7$. The comparison is shown in Figure\,\ref{fig:fa}(d) and the corresponding $h(2)$ values are quoted in Table \ref{tab1}. The superiority of any one of the methods over the other, in connection with real data, is not yet thoroughly studied. However, there exist some evidences where the MFDMA analysis method is found to be more useful than the other one \cite{Gu10,Ruan11,Shao12}. In MFDMA analysis the parameter $\theta=0.7$ implies that a measurement in the records is to be detrended by a window composed of 30\% backward and 70\% forward memories. In reality forward memory of a time series may not be a convenient concept. But one may think it in this way: any measurement $x_i$ in a time series which already carries a past memory/persistence of about 30\% might influence the $x_{i+1}$th measurement by at best 70\%. In that sense, the global temperature anomaly time series is highly long-range correlated and the correlation itself might be the main source of the observed multifractality.

\section{Conclusions}
\label{Conclusions}
In this article we present the multifractal detrended moving average analysis of global monthly mean temperature anomaly time series over the period of 1850--2012. Various observable related to (multi)fractals, namely the generalised Hurst exponent $h(q)$, the multifractal exponent $\tau(q)$ and the multifractal singularity spectra are calculated for the temperature anomaly records. We find that the global monthly mean temperature records are of multifractal nature and the main source of it is the long-range correlation in the measurements. The multifractal signature of the time series is also obtained from autocorrelation function analysis. The results of this analysis are found to be comparable with that of the MFDFA (first order) method provided the detrending moving window for an arbitrary measurement is constructed out of 30\% backward and 70\% forward memories with respect to the measurement. Till date MFDMA is not widely applied to analyze time series data of different variants. Therefore, a systematic comparative study between MFDMA analysis and other known methods as well as various multifractal models would be a highly encouraging exercise which might help us to visualize the predictability and hence applicability of the MFDMA method in time series analysis.

\section*{Acknowledgement}
The author thanks Soumya Sarkar of the department of physics, NBU for drafting some of the figures shown in this article. Many constructive suggestions of the reviewers are gratefully acknowledged.

\section*{References}
\bibliographystyle{plain}

\begin{thebibliography}{99}
\bibitem{Mand1} Mandelbrot B B, van Ness J W, 1968 {\it SIAM Review} {\bf 10}  422
\bibitem{Mand2} Mandelbrot B B, Wallis J R, 1969 {\it Water Resour. Res.} {\bf 5} 321

\bibitem{Bara91} Barab\'asi A-L, Vicsek T, 1991 {\it Phys. Rev.} A {\bf 44} 2730
\bibitem{Peit92} Peitgen H -O, Jurgens H, Saupe D, 1992 {\it Chaos and Fractals} (Springer, New York) (Appendix B).
\bibitem{Barcy01} Bacry E, Delour J, Muzy J F, 2001 {\it Phys. Rev.} E {\bf 64} 026103

\bibitem{Muzy91} Muzy J F, Bacry E, Arneodo A, 1991 {\it Phys. Rev. Lett.} {\bf 67} 3515
\bibitem{Muzy94} Muzy J F, Bacry E, Arneodo A, 1994 {\it Int. J. Bifurcat. Chaos} {\bf 4} 245
\bibitem{Arne95} Arneodo A, Bacry E, Graves P V, Muzy J F, 1995 {\it Phys. Rev. Lett.} {\bf 74} 3293
\bibitem{Ivan99} Ivanov P ch, Amaral L A N, Goldberger A L, Havlin S, Rosenblum M G, Struzik Z R,  Stanley H E, 1999 {\it Nature} {\bf 399} 461
\bibitem{Amar01} Amaral L A N, Ivanov P Ch, Aoyagi N, Hidaka I, Tomono S, Goldberger A L, Stanley H E,  Yamamoto Y, 2001 {\it Phys. Rev. Lett.} {\bf 86} 6026
\bibitem{Silc01} Silchenko A, Hu C K, 2001 {\it Phys. Rev.} E {\bf 63} 041105
\bibitem{Kant02} Kantelhardt J W, Zschiegner S A, Koscielny-Bunde E, Havlin S, Bunde A, Stanley H E, 2002  {\it Physica} A {\bf 316} 87

\bibitem{Peng94} Peng C -K, Buldyrev S V, Havlin S, Simons M, Stanley H E, Goldberger A L, 1994 {\it Phys. Rev.} E {\bf 49} 1685
\bibitem{Ossa94} Ossadnik S M, Buldyrev S B, Goldberger A L, Havlin S, Mantegna R N,  Peng C -K, Simons M, Stanley H E, 1994 {\it Biophys. J.} {\bf 67} 64

\bibitem{Eichner03} Eichner J F, Koscielny-Bunde E, Bunde A, Havlin S, Schellnhuber H J, 2003 {\it Phys. Rev.} E {\bf 68} 046133
\bibitem{Maraun04} Maraun D, Rust H W, Timmer J, 2004 {\it Nonlinear Proc. Geophys.} {\bf 11} 495

\bibitem{Varotsos05} Varotsos C, 2005 {\it International Journal of Remote Sensing} {\bf 26} 3333

\bibitem{Varo13} Varotsos C A, Efstathiou M N, Cracknell A P, 2013 {\it Atmos. Chem. Phys.} {\bf 13}  5243

\bibitem{Gu10} Gu G -F, Zhou W X, 2010 {\it Phys. Rev.} E {\bf 82} 011136
\bibitem{Ales02} Alessio E,  Carbone A, Castelli G, Frappietro V, 2002 {\it Eur. Phys. J.} B {\bf 27} 197

\bibitem{Schu11} Schumann A Y, Kantelhardt J W, 2011 {\it Physica} A {\bf 390} 2637
\bibitem{Wang11} Wang Y, Wu C, Pan Z, 2011 {\it Physica} A {\bf 390} 3512
\bibitem{Ruan11} Ruan Y -P, Zhou W-X, 2011 {\it Physica} A {\bf 390} 1646
\bibitem{Shao12} Shao Y H, Gu G F, Jiang Z Q {\it et al.}, 2012 {\it Sci. Rep.} {\bf 2} 835
\bibitem{Zhou12} Zhou W X, 2012 {\it Chaos, Solitons \& Fractals} {\bf 45} 147
\bibitem{Wang14} Wang F, Wang L, Zou R -B, 2014 {\it Chaos} {\bf 24} 033127

\bibitem{Kosc98} Koscielny-Bunde E, Bunde A, Havlin S, Roman H R, Goldreich Y, Schellnhuber H J, 1998  {\it Phys. Rev. Lett.} {\bf 81} 729
\bibitem{Weber01} Weber R O, Talkner P, 2001 {\it J. Geophys. Res.} {\bf 106} 20131
\bibitem{Efst13} Efstathiou M N, Varotsos C A, 2013 {\it Meteorol. Appl.} {\bf 20} 72
\bibitem{Mali14a} Mali P, 2015 {\it Theor. Appl. Climatol.} {\bf 121} 641
\bibitem{Kant03} Kantelhardt J W, Rybski D, Zschiegner S A, Braun P,  Koscielny-Bunde E, Livina V, Havlin S, Bunde A, 2003 {\it Physica} A {\bf 330} 240

\bibitem{Data}  Jones P D,  Parker D E, Osborn T J,  Briffa K R, 2013 {\it Trends: A Compendium of Data on Global Change, Carbon Dioxide Information Analysis Center} (Oak Ridge National Laboratory, U.S. Department of Energy, Oak Ridge. Tenn USA) DOI: 10.3334/CDIAC/cli.002


\bibitem{Chianka05} Chianca C V, Ticona A, Penna T J P, 2005 {\it Physica} A {\bf 357} 447–54

\bibitem{Mova06} Movahed M S, Jafari G R, Ghasemi F, Rahvar S, Tabar M R R, 2006 {\it J. Stat. Mech.} P02003 

\bibitem{Halsey86} Halsey T C, Jensen M H,  Kadanoff L P, Procaccia I, Shraiman B I, 1986 {\it Phys. Rev.} A {\bf 33} 1141
\bibitem{Pei92} Peitgen H -O, Jurgens H, Saupe D, 1992 {\it Chaos and Fractals} (Springer, New York)  Appendix B


\bibitem{Bohr87a} Bohr T,  Jensen H M, 1987 {\it Phys. Rev.} A {\bf 36} 321
\bibitem{Bohr87b} Bohr T,  Rand D, 1987 {\it Physica} A {\bf 25} 387

\bibitem{Xu05} Xu L, Ivanov P Ch, Hu K, Chen Z, Carbone A, Stanley H E, 2005 {\it Phys. Rev.} E {\bf 71} 051101. 
\bibitem{Alfio}  Quarteroni A, Sacco R, Saleri F, 2007 {\it Numerical Mathematics} (Springer Berlin Heidelberg) p 20

\bibitem{Ulenbeck}  Uhlenbeck G E, Ornstein L S, 1930 {\it Phys. Rev.} {\bf 36} 823
\bibitem{Thomas} Schreiber T, Schmitz A, 2000 {\it Physica} D {\bf 142} 346

\bibitem{Theiler} Theiler J, Eubank S, Longtin A, Galdrikian B, Farmer J D, 1992 {\it Physica} D {\bf 58} 77   
\bibitem{Alfi06}  Alfi V, Coccetti F, Petri A,  Pietronero L, 2007 {\it Euro. Phys. J.} B {\bf 55} 135
\bibitem{Bashan08}Bashan A, Bartsch R,  Kantelhardt J W, Havlin S, 2008 {\it Physica} A {\bf 387} 5080

\bibitem{Avnir98} Avnir D, Birham O, Lindar D, Malcai O, 1998 {\it Science} {\bf 279} 39




\bibitem{Gires} Gires A, Tchiguirinskaia I, Schertzer D, Lovejoy S, 2013 {\it Nonlin. Proc. Geophys.} {\bf 20} 343 
\bibitem{Yu14} Yu Z -G, Leung Y, Chen Y D, Zhang Q, Anh V, Zhou Y, 2014 {\it Physica} A {\bf 405} 193

\bibitem{Mali14} Mali P, Mukhopadhyay A, 2014 {\it Physica} A {\bf 413} 361
\bibitem{Mali15} Mali P, Sarkar S, Ghosh S, Mukhopadhyay A, Singh G, 2015 {\it Physica} A {\bf 424} 25



\end{thebibliography}

\end{document}